\documentclass[9pt,conference]{IEEEtran}
\IEEEoverridecommandlockouts

\usepackage[sort]{cite}
\usepackage{amsmath,amssymb,amsfonts}
\usepackage{algorithmic}
\usepackage{textcomp}
\PassOptionsToPackage{table,xcdraw,dvipsnames}{xcolor}
\usepackage{xcolor}
\usepackage{pbalance}
\PassOptionsToPackage{hyphens}{url}
\PassOptionsToPackage{
  colorlinks=true,
  linkcolor=blue,
  citecolor=green,
  urlcolor=cyan,
  hidelinks,
  bookmarks=false
}{hyperref}
\usepackage{hyperref}
\usepackage[justification=centering]{caption}
\usepackage{subfigure} 
\usepackage{booktabs} 
\usepackage{multirow}  
\usepackage{siunitx}
\usepackage{array}

\newcolumntype{M}[1]{>{\centering\arraybackslash}m{#1}} 
\usepackage{waspaa25}

\usepackage{acronym}
\acrodef{SRO}{sample rate offset}
\acrodef{AIR}{acoustic impulse response}
\acrodef{RTF}{relative transfer function}
\acrodef{ppm}{parts-per-million}
\acrodef{STFT}{short-time Fourier transform}
\acrodef{ATF}{acoustic transfer function}
\acrodef{DWACD}{dynamic weighted average coherence drift}
\acrodef{GCC}{generalized cross-correlation}
\acrodef{LCMV}{linearly constrained minimum variance}
\acrodef{PSD}{power spectral density}
\acrodef{ITD}{interaural time difference}
\acrodef{ILD}{interaural level difference}
\acrodef{IC}{interaural coherence}
\acrodef{PTP}{Precision Time Protocol}
\acrodef{NTP}{Network Time Protocol}
\acrodef{MDO}{media device orchestration}
\acrodef{MUSHRA}{Multiple Stimuli with Hidden Reference and Anchor}

\title{Stereo Reproduction in the Presence of Sample Rate Offsets}

\name{Srikanth Korse$^{1,\dag}$,\thanks{*A joint institution of the Friedrich-Alexander-Universit\"{a}t Erlangen-N\"{u}rnberg (FAU) and Fraunhofer IIS, Germany.\\ \dag The author completed this work while at the International Audio Laboratories Erlangen and is now with Nokia Technologies.}
      Andreas Walther$^{2}$,
      and Emanu\"{e}l\,A.\,P.\, Habets$^{1,2}$
      }
\address{$^{1}$International Audio Laboratories Erlangen*, Erlangen, Germany \;
$^{2}$Fraunhofer IIS, Erlangen, Germany\\
}

\begin{document}

\maketitle

\begin{abstract}

One of the main challenges in synchronizing wirelessly connected loudspeakers for spatial audio reproduction is clock skew. Clock skew arises from \acp{SRO} between the loudspeakers, caused by the use of independent device clocks.
While network-based protocols like Precision Time Protocol (PTP) and Network Time Protocol (NTP) are explored, the impact of \acp{SRO} on spatial audio reproduction and its perceptual consequences remains underexplored. We propose an audio-domain \ac{SRO} compensation method using spatial filtering to isolate loudspeaker contributions. These filtered signals, along with the original playback signal, are used to estimate the \acp{SRO}, and their influence is compensated for before spatial audio reproduction. We evaluate the effect of the compensation method in a subjective listening test. The results of these tests, as well as objective metrics, demonstrate that the proposed method 
mitigates the perceptual degradation introduced by \acp{SRO} by preserving the binaural cues.

\end{abstract}

\begin{IEEEkeywords}
Spatial Audio, Spatial Audio Reproduction, Sample Rate Offset, Synchronization, Spatial Filtering
\end{IEEEkeywords}

\acresetall

\section{Introduction}


Spatial audio capturing and reproduction enable a wide range of applications in entertainment and teleconferencing \cite{Rumsey_2001, Roginska_2017}. On the playback side, spatial audio reproduction aims to recreate the captured complex acoustic environments - or construct completely new ones - allowing listeners to perceive localized sounds from dedicated positions in space and to experience a sense of envelopment.


A key challenge in spatial audio reproduction over wirelessly connected loudspeakers is the synchronization of the individual playback devices. One important factor in this regard is clock skew caused by \acp{SRO} between devices with individual clocks. \acp{SRO} cause time-varying misalignment of playback signals, leading to degradation of binaural cues. Existing clock synchronization methods rely on network-based protocols such as \ac{PTP}\,\cite{ieee1588-2008} and \ac{NTP}\,\cite{rfc5905}, which aim to align device clocks\,\cite{Culbert_2006,Lauri2015}. However, to the best of our knowledge, the impact of the presence of \acp{SRO} on the listener's perception in spatial audio reproduction has not been systematically studied.



While \ac{SRO} estimation using audio-domain observations has been extensively explored in wireless sensor networks\,\cite{SMG_BSRO_2012,wang_CMSRO_WASN_2016,schmalenstroeer_multi-stage_2017,chinaev_online_2021,gburrek_synchronization_2022} and acoustic echo cancellation\,\cite{pawig_adaptive_2010,abe_frequency_2014,helwani_clock_2022,korse_2024}, no prior work has investigated this in the context of spatial audio reproduction specifically.

In this work, we examine the effects of \acp{SRO} in a stereo loudspeaker scenario. Further, we propose an audio-domain \ac{SRO} compensation method that applies spatial filtering\,\cite{VanVeen_1988} to isolate loudspeaker contributions.  The isolated loudspeaker signals are then used along with the original playback signals to estimate the \acp{SRO} using the~\ac{DWACD} algorithm~\cite{gburrek_synchronization_2022}. We then compensate for the effect of \acp{SRO} before the spatial rendering step, thus eliminating the need for explicit clock synchronization. Evaluation using objective metrics and subjective listening tests demonstrates that the proposed method preserves binaural cues and mitigates perceptual degradation.

\section{ Stereo Reproduction amid \acp{SRO} } 

\begin{figure}[t]
  \centering
  \includegraphics[width=0.65\columnwidth]{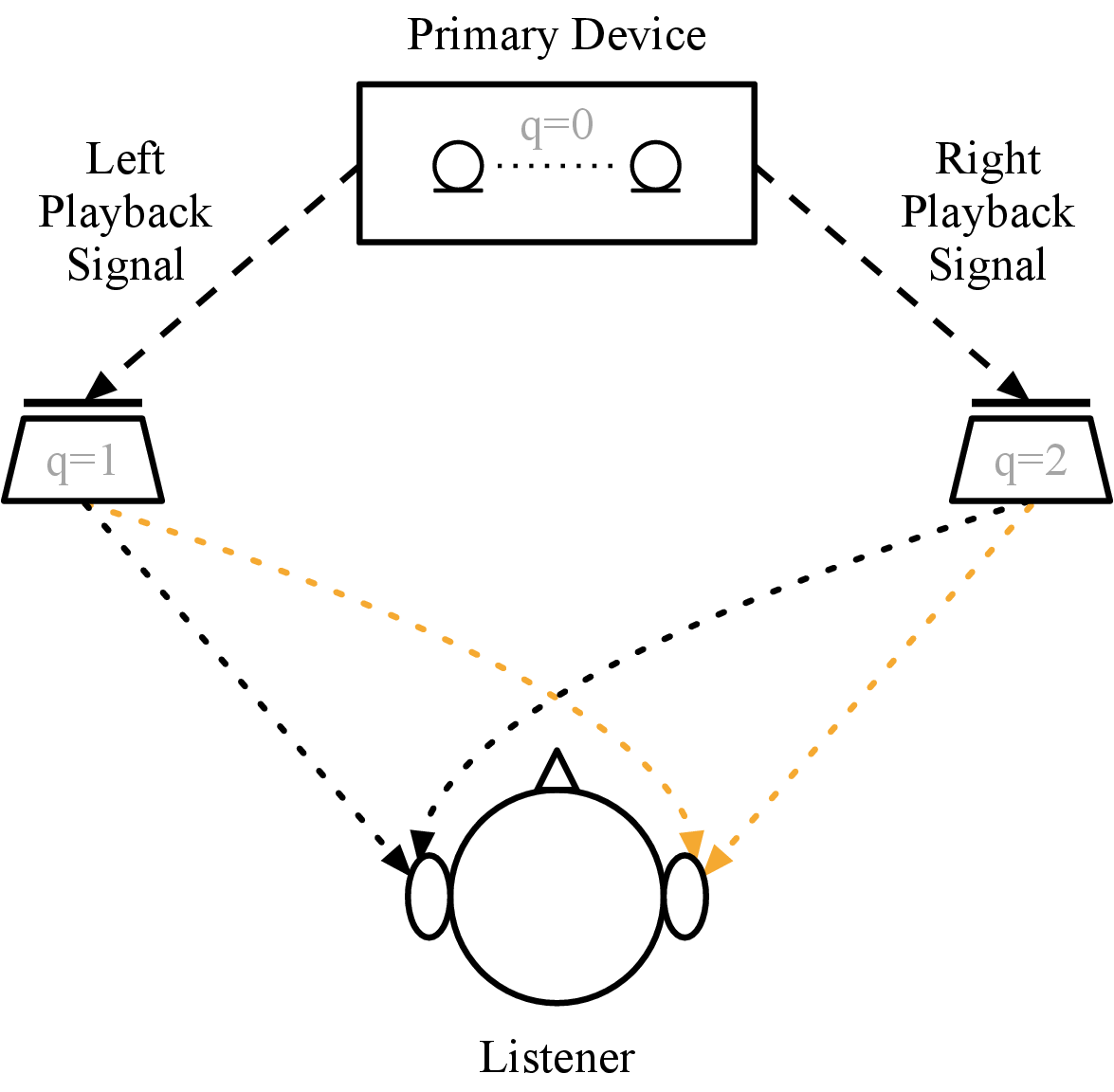}
\caption{Stereo reproduction using two devices, each consisting only of a loudspeaker ($q \in {1,2}$), wirelessly connected to a primary device that consists only of a microphone array (q=0). }
\label{fig:spatial_reproduction}
\end{figure} 

Consider a room with three devices and a listener, as shown in Fig.\,\ref{fig:spatial_reproduction}. Without loss of generality, we define the first device as the primary device ($q = 0$), which transmits playback signals at a sampling rate $f_s$ to auxiliary devices ($q \in {1,2}$), each containing a single loudspeaker. We assume that the primary device contains only a microphone array. Under the above assumptions, the binaural signal at the ears of the listener $b_{i}$, where $i \in \{L,R\}$ can be described as
\begin{equation}
\label{equ:listener_sig_TD}
b_{i}\left[n\right] = \sum_{q=1}^{2} h_{i,q}\left[n\right] * x_{q}\left[n\right],
\end{equation}
where $n$ is the sample index, $q$ is the index of the auxiliary device, $x_q\left[n\right]$ is the  playback signal, $h_{i,q}[n]$ are the \acp{AIR} between the loudspeakers of the $q\text{-th}$ auxiliary device and the ears of the listener. Assuming that the playback signals are synchronized, i.e., the auxiliary devices play at the same sampling rate $f_s$, \eqref{equ:listener_sig_TD} can be described in the time-frequency domain as
\begin{equation}
\label{equ:listener_sig_FD}
B_{i}\left[k,l\right] = \sum_{q=1}^{2} H_{i,q}\left[k,l\right] \, X_{q}\left[k,l\right],
\end{equation}
where $B_{i}\left[k,l\right]$, $X_{q}\left[k,l\right]$ and $H_{i,q}\left[k,l\right]$ with frequency index~$k$ and frame index~$l$ represent the time-frequency domain counterparts of $b_i[n]$, $x_{q}\left[n\right]$ and $h_{i,q}\left[n\right]$, respectively. However, when the sampling rates of the auxiliary devices differ due to the presence of \acp{SRO}, the playback signals are not synchronized. In such a scenario, the binaural signal at the ears of the listener in the frequency domain with sufficiently large window size $N_\text{w}$ and hop size $N_\text{h}$ can be written as~\cite{SMG_BSRO_2012}
\begin{equation}
\label{equ:listener_sig_FD_SRO}
B_{i}\left[k,l\right] = \sum_{q=1}^{2} H_{i,q}\left[k,l\right] \, \Lambda_{q}\left[k,l\right]\, X_{q}\left[k,l\right],
\end{equation}
where $\Lambda_{q}\left[k,l\right] = \text{exp}\left(\frac{-j 2\pi k}{N_\text{w}}\left(\frac{l\,N_\text{h}\,{\epsilon}_{q}}{f_{s}}\right)\right)$, ${\epsilon}_{q}$ is the \ac{SRO} defined in \ac{ppm}. Note that \eqref{equ:listener_sig_FD_SRO} holds only if $\frac{l\,N_\text{h}\,{\epsilon}_{q}}{f_{s}} \ll N_\text{w}$~\cite{wang_CMSRO_WASN_2016}. The \ac{SRO} term explains the difference between the sampling rate of the playback signals $f_s$ and the playback sampling rate $f_q$ and is given by
\begin{equation}
\label{equ:sro_equ}
f_{q} = (1+\epsilon_q)\,f_s.
\end{equation}
In the following, we assume $\epsilon_q$ is constant and ignore the effect of delay, coding, and frame errors that commonly occur during the transmission of playback signals from the primary device to the auxiliary devices. 

In this study, we evaluate the effect of $\Lambda_{q}\left[k,l\right]$ on spatial perception from the listener's perspective. In addition, we propose a solution to nullify the effect of $\Lambda_{q}\left[k,l\right]$. This involves first estimating the \ac{SRO} ${\epsilon}_{q}$ and resampling the playback signal $X_{q}\left[k,l\right]$ to $\bar{X}_{q}\left[k,l\right]$ where $\bar{X}_{q}\left[k,l\right] = X_{q}\left[k,l\right]\,\bar{\Lambda}_{q}\left[k,l\right]$ such that $\bar{\Lambda}_{q}\left[k,l\right]\,\Lambda_{q}\left[k,l\right] = 1$ and $\bar{\Lambda}_{q}\left[k,l\right] = \text{exp}\left(\frac{j2\pi k}{N_\text{w}}\left(\frac{l\,N_\text{h}\,{\epsilon}_{q}}{f_{s}}\right)\right)$.

\section{\ac{SRO}-Compensated Stereo Reproduction} 

\begin{figure}[t]
  \centering
  \includegraphics[width=\columnwidth]{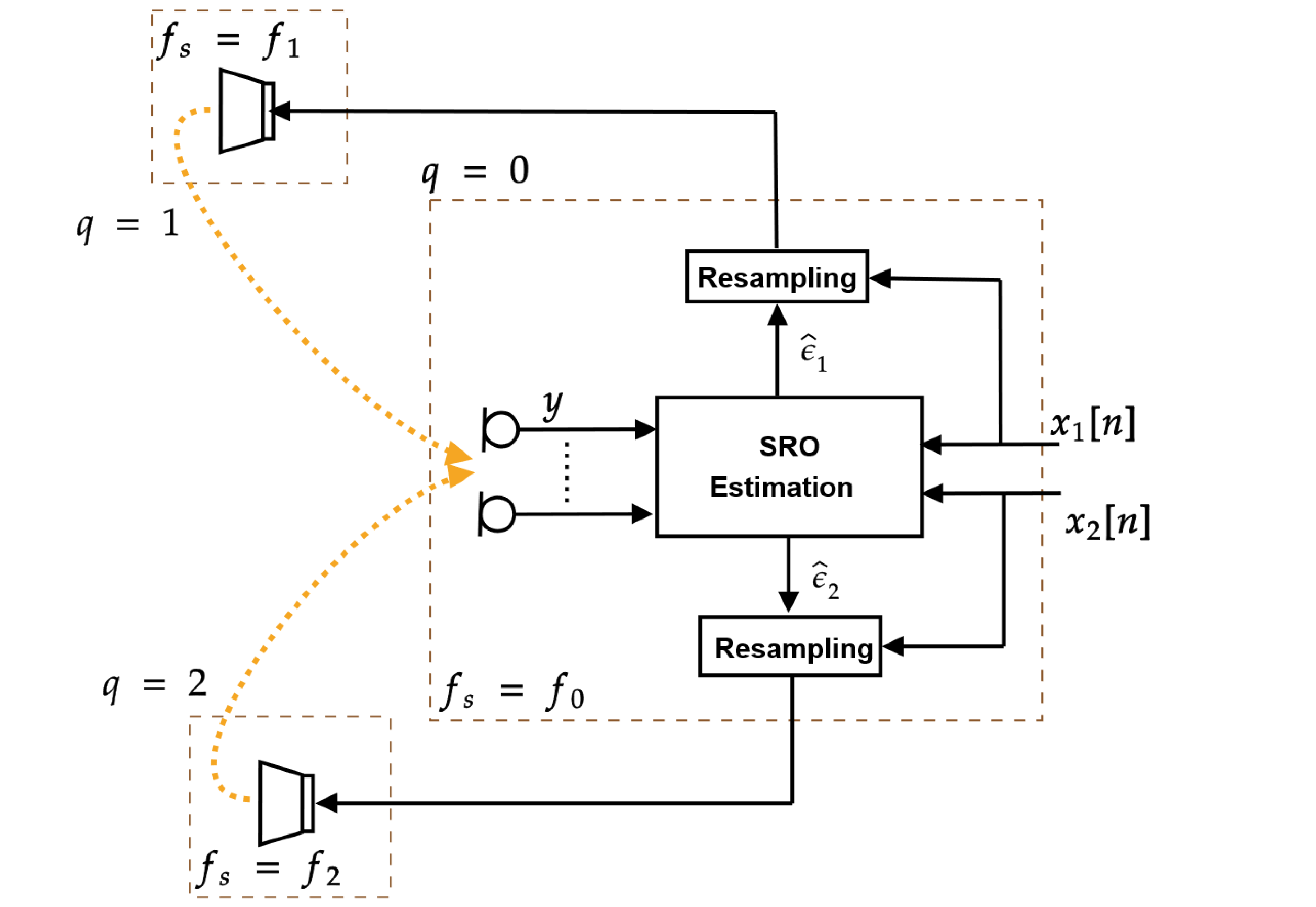}
\caption{\ac{SRO}-Compensated Stereo Reproduction.} 
\label{fig:sro_compensated_spatial_reproduction}
\end{figure}

Our proposed \ac{SRO}-compensated spatial reproduction system is as shown in Fig.\,\ref{fig:sro_compensated_spatial_reproduction}. In this setup, the primary device, which includes a microphone array, is responsible for estimating the \ac{SRO} and resampling the playback signal before transmitting it to the auxiliary devices. The primary device uses a spatial filter\,\cite{VanVeen_1988,VanTrees2002} to extract each individual contribution of the loudspeaker signals. The spatial filter output and the playback signal are then used to estimate the \ac{SRO}. The playback signal is subsequently resampled to compensate for the effect of \ac{SRO} before being transmitted to the auxiliary devices. 

\subsection{Signal Model}
The microphone signals of the primary device can be defined as
\begin{equation}
\label{equ:mic_sig_TD}
y_{0,m}\left[n\,T_{0}\right] = \sum_{q=1}^{2} h_{0,q,m}[n\,T_{0}] * x_{q}\left[n\,T_{0}\right] + v_{0,m}\left[n\,T_{0}\right],
\end{equation}
where $n$ is the sample index, $T_0 = \frac{1}{f_0}$ is the sampling period of the primary device, $m \in \{1,2,\ldots,M\}$ is the microphone index, $q$ is the index of the auxiliary device, $x_q\left[n\,T_{0}\right]$ denotes the playback signal sent to the $q\text{-th}$ loudspeaker, $v_{0,m}\left[n\,T_{0}\right]$ is the contribution of sensor noise, $h_{0,q,m}[n\,T_{0}]$ is the \ac{AIR} between the loudspeaker of the $q\text{-th}$ auxiliary device and the $m\text{-th}$ microphone of the primary device. Since the loudspeaker of the $q\text{-th}$ auxiliary device converts the digital signal $x_q\left[n\,T_s\right]$ that is transmitted from the primary device into its analog counterpart at a sampling rate of $f_q$ and the microphone of the primary device converts the analog signal back to a digital signal at a sampling rate of $f_0$, we can define the relation between $f_s$ and $f_0$ in terms of \ac{SRO} between the devices $\epsilon_{q}$ and $\epsilon_{0}$ as
\begin{equation}
\label{equ:relation_frequencies_SRO}
f_0 = (1+\epsilon_{q})\,(1+\epsilon_{0})\,f_s,
\end{equation}
where $|\epsilon_{q}| \ll 1$ and $|\epsilon_{0}| \ll 1$ are usually expressed in~\ac{ppm}. Since the terms $\epsilon_{q}$ and $\epsilon_{0}$ are very small, the term $\epsilon_{q}*\epsilon_{0}$ can be ignored. This further simplifies \eqref{equ:relation_frequencies_SRO} as
\begin{equation}
\label{equ:relation_frequencies_SRO_1}
f_0 = (1+\epsilon_{q}+\epsilon_{0})\,f_s = (1+\bar{\epsilon}_{q})\,f_s.
\end{equation}

Substituting \eqref{equ:relation_frequencies_SRO_1} in \eqref{equ:mic_sig_TD}, applying Taylor series approximation to the term $x_{q}\left[\frac{n}{(1+\bar{\epsilon}_{q})\,f_s}\right]$~\cite{SMG_BSRO_2012},~\eqref{equ:mic_sig_TD} can be approximated in the time-frequency domain with sufficiently long window size $N_\text{w}$ and hop size $N_\text{h}$ as
\begin{multline}
\label{equ:mic_sig_FD}
Y_{0,m}\left[ k, l \right] = \sum_{q=1}^{2} H_{0,q,m}\left[ k, l \right] ~\Lambda_{q}\left[ k, l \right] ~X_{q}\left[ k, l \right]   \\
 + V_{0,m}\left[ k, l \right],
\end{multline}
where $\Lambda_{q}\left[ k, l \right]  = \text{exp}\left(\frac{-j2\pi k}{N_\text{w}}\left(\frac{l\,N_\text{h}\,\bar{\epsilon}_{q}}{f_{s}}\right)\right)$, $X_{q}\left[ k, l \right]$, $H_{0,q,m}\left[ k, l \right]$ and $V_{0,m}\left[ k, l \right]$ with frequency index~$k$ and frame index~$l$ are the frequency-domain representation of $x_{q}\left[n\,T_{s}\right]$, $h_{0,q,m}[n\,T_0]$ and $v_{0,m}\left[n\,T_{0}\right]$, respectively. Similarly to \eqref{equ:listener_sig_FD_SRO}, \eqref{equ:mic_sig_FD} also holds only if the condition $\frac{l\,N_\text{h}\,\bar{\epsilon}_{q}}{f_{s}} \ll N_\text{w}$ is satisfied~\cite{wang_CMSRO_WASN_2016}. Since the primary device estimates the term $\bar{\epsilon}_{q}$, in the current work, we assume that $\epsilon_0$ is known a priori to estimate $\epsilon_q$ from $\bar{\epsilon}_{q}$. 

\subsection{Playback Signal-Assisted Spatial Filtering} 
In vector notation, \eqref{equ:mic_sig_FD} can be written as
\begin{equation}
\label{equ:mic_sig_FD_vecNotation}
\mathbf{y} = \mathbf{H}\,\mathbf{\Lambda}\,\mathbf{x} + \mathbf{v},
\end{equation}
where the microphone signal $\mathbf{y}$, playback signal $\mathbf{x}$, \ac{ATF} $\mathbf{H}$, \ac{SRO} contribution $\mathbf{\Lambda}$ and sensor noise $\mathbf{v}$ are defined as
\begin{equation}
\label{equ:mic_sig_vec_notation}
\mathbf{y} = [Y_{0,0}\left[ k, l \right],Y_{0,1}\left[ k, l \right] \cdots Y_{0,M-1}\left[ k, l \right]]^T \in R^{M \times 1},
\end{equation}
\begin{equation}
\label{equ:playback_sig_notation}
\mathbf{x} = [X_{1}\left[ k, l \right],X_{2}\left[ k, l \right]]^T \in R^{2 \times 1},
\end{equation}
\begin{equation}
\label{equ:ATF_vec_notation}
\mathbf{H} = [\mathbf{h}_{0,1},\mathbf{h}_{0,2}] \in R^{M \times 2},
\end{equation}
\begin{equation}
\label{equ:SRO_contribution_vec_notation}
\mathbf{\Lambda} = \text{diag}[\Lambda_{1}\left[ k, l \right],\Lambda_{2}\left[ k, l \right]] \in R^{2 \times 2},
\end{equation}
\begin{equation}
\label{equ:sensor_noise}
\mathbf{v} = [V_{0,0}\left[ k, l \right],V_{0,1}\left[ k, l \right] \cdots V_{0,M-1}\left[ k, l \right]]^T \in R^{M \times 1},
\end{equation}
where $\mathbf{h}_{0,q} = [H_{0,q,0}\left[ k, l \right],H_{0,q,1}\left[ k, l \right] \cdots H_{0,q,M-1}\left[ k, l \right]]^T$. 
Since we aim to robustly estimate $\bar{\epsilon}_{1}$ and $\bar{\epsilon}_{2}$, we need to extract individual terms $\mathbf{h}_{0,1}\,\Lambda_{1}\left[ k, l \right]\,X_{1}\left[ k, l \right]$ and $\mathbf{h}_{0,2}\,\Lambda_{2}\left[ k, l \right]\,X_{2}\left[ k, l \right]$ from the microphone signal $\mathbf{y}$. In this study, we employ a well-known \ac{LCMV} beamformer\,\cite{VanVeen_1988} to extract individual contributions from the microphone signal. The $q{\text{-th}}$ output of the beamformer $\widehat{Z}_q\left[ k, l \right]$ can be described as
\begin{equation}
\label{equ:beamformer_ouput}
\widehat{Z}_q\left[ k, l \right] = \mathbf{w}^{H}_{q}\left[ k, l \right]\,\mathbf{y},
\end{equation}
where $\mathbf{w}_{q}\left[ k, l \right]$ contains the beamformer weights computed by treating the $q\text{-th}$ loudspeaker as the "source of interest" and the other loudspeaker as the interferer, and superscript $(\cdot)^H$ denotes the Hermitian. Ideally, $\widehat{Z}_q\left[ k, l \right] \approx \mathbf{h}_{0,q}\,\Lambda_{q}\left[ k, l \right]\,X_{q}\left[ k, l \right] + \mathbf{v}$. The \ac{LCMV} beamformer $\mathbf{w}_{q}\left[ k, l \right]$ is computed by solving the optimization: problem\,\cite{Frost1972}
\begin{multline}
\label{equ:lcmv_optim_problem}
\mathbf{w}_{q}\left( k, l \right) = \arg \min_{\mathbf{w}_{q}} \mathbf{w}_{q}^{H}\,\mathbf{\Phi}_{v}\,\mathbf{w_{q}} \\
\text{subject to}~\mathbf{w}_{q}^{H}\,\mathbf{A}\left[ k, l \right] = \mathbf{g}_q,
\end{multline}
where the gain $\mathbf{g}_q$ is the $q\text{-th}$ column of the identity matrix $\mathbf{I} \in \mathbb{R}^{2 \times 2}$, $\mathbf{A}\left[ k, l \right] = [\mathbf{a}_{0}, \mathbf{a}_{1}] \in R^{M \times 2}$ is the \ac{RTF} matrix, $\mathbf{a}_{q} = \frac{\mathbf{h}_{0,q}}{H_{0,q,0}}$ is the \ac{RTF} vector under the assumption that $0\text{-th}$ microphone is the reference microphone.
The solution to the above optimization problem is given by
\begin{equation}
\label{equ:lcmv_bf}
\mathbf{w}_{q}\left[ k, l \right] = \mathbf{\Phi}_{v}^{-1}\,\mathbf{A}\left[ k, l \right]\,\left[ \mathbf{A}^{H}\left[ k, l \right]\,\mathbf{\Phi}_{v}^{-1}\,\mathbf{A}\left[ k, l \right]\right]^{-1} \,
\mathbf{g}_{q}, 
\end{equation}
where $\mathbf{\Phi}_{v}$ is the noise \ac{PSD} matrix. The inverse term in \eqref{equ:lcmv_bf} exists only if it satisfies the following criteria\,\cite{VanVeen_1988, VanTrees2002, chakrabarty2016numerical}: i) $\mathbf{\Phi}_{v}$ is full-rank, ii) $\mathbf{A}\left[ k, l \right]$ has linearly independent columns. We set $\mathbf{\Phi}_{v} = \mathbf{I}$ in our implementation, which satisfies the first criterion. However, since we cannot guarantee that the second criterion is fulfilled, we use a simple regularization method called diagonal loading, where a small term $\alpha\,\mathbf{I}$ is added to the inverse term\,\cite{bishop2006pattern, chakrabarty2016numerical} to ensure numerical stability. With the above, the solution to the \ac{LCMV} beamformer is given by
\begin{equation}
\label{equ:lcmv_bf_v1}
\mathbf{w}_{q}\left[ k, l \right] = \mathbf{A}\left[ k, l \right]\,\left[ \mathbf{A}^{H}\left[ k, l \right]\,\mathbf{A}\left[ k, l \right] + \alpha\,\mathbf{I}\right]^{-1}\,\mathbf{g}_{q}, 
\end{equation}
where $\alpha$ is a small constant. 

Accurate estimation of the \ac{RTF} matrix $\mathbf{A}\left[ k, l \right]$ is essential for the computation of the beamformer weights defined in \eqref{equ:lcmv_bf_v1}. There exist many methods to estimate \ac{RTF}, among them, minimum distortion-based estimator\,\cite{Chen2008} and subspace-based estimators\,\cite{Markovich2009} can be considered as state-of-the-art estimators. These estimators work well for one source; however, in the case of multiple sources, these estimators require periods where only one source is active. The authors of\,\cite{Taseska_2015} proposed a time-varying \ac{RTF} estimator per time-frequency bin corresponding to the dominant source at that bin. Since the \ac{RTF} estimation is not the main focus of the paper, in the current study, we use an oracle \ac{RTF}, which is computed as
\begin{equation}
    \label{equ:oracle_RTF}
    \mathbf{a}_q\left[ k, l \right] = \frac{\mathbf{\bar{\mathbf{\Phi}}}_q}{\mathbf{e}^T\mathbf{\bar{\mathbf{\Phi}}}_q}, 
\end{equation}
where $\mathbf{e}=[1,0,\cdots0]^T$ is the selection vector, $\mathbf{\bar{\mathbf{\Phi}}}$ is the PSD matrix defined as
\begin{equation}
    \label{equ:cross_covariance_marix}
    \mathbf{\bar{\mathbf{\Phi}}}_q = \mathbf{E}\{\mathbf{\bar{z}}_q\,X^{*}_q\}, 
\end{equation}
where $(\cdot)^*$ denotes the complex conjugate and $\mathbf{\bar{z}}_q $ denotes the non-\ac{SRO} compensated microphone signal containing only the \ac{SRO} affected playback signal $X_q$ and is given by:
\begin{equation}
  \label{equ:mic_sig_oracle}
\mathbf{\bar{z}}_q = \mathbf{h}_{0,q}~\Lambda_{q}\left[ k, l \right] ~X_{q}\left[ k, l \right] + \mathbf{v}.  
\end{equation}
In this study, we assume a initialization phase where only $q\text{-th}$ loudspeaker is active. 

\subsection{SRO Estimation}

\begin{figure*}[t]
  \centering
  \subfigure[ITD]{
    \includegraphics[width=0.485\linewidth]{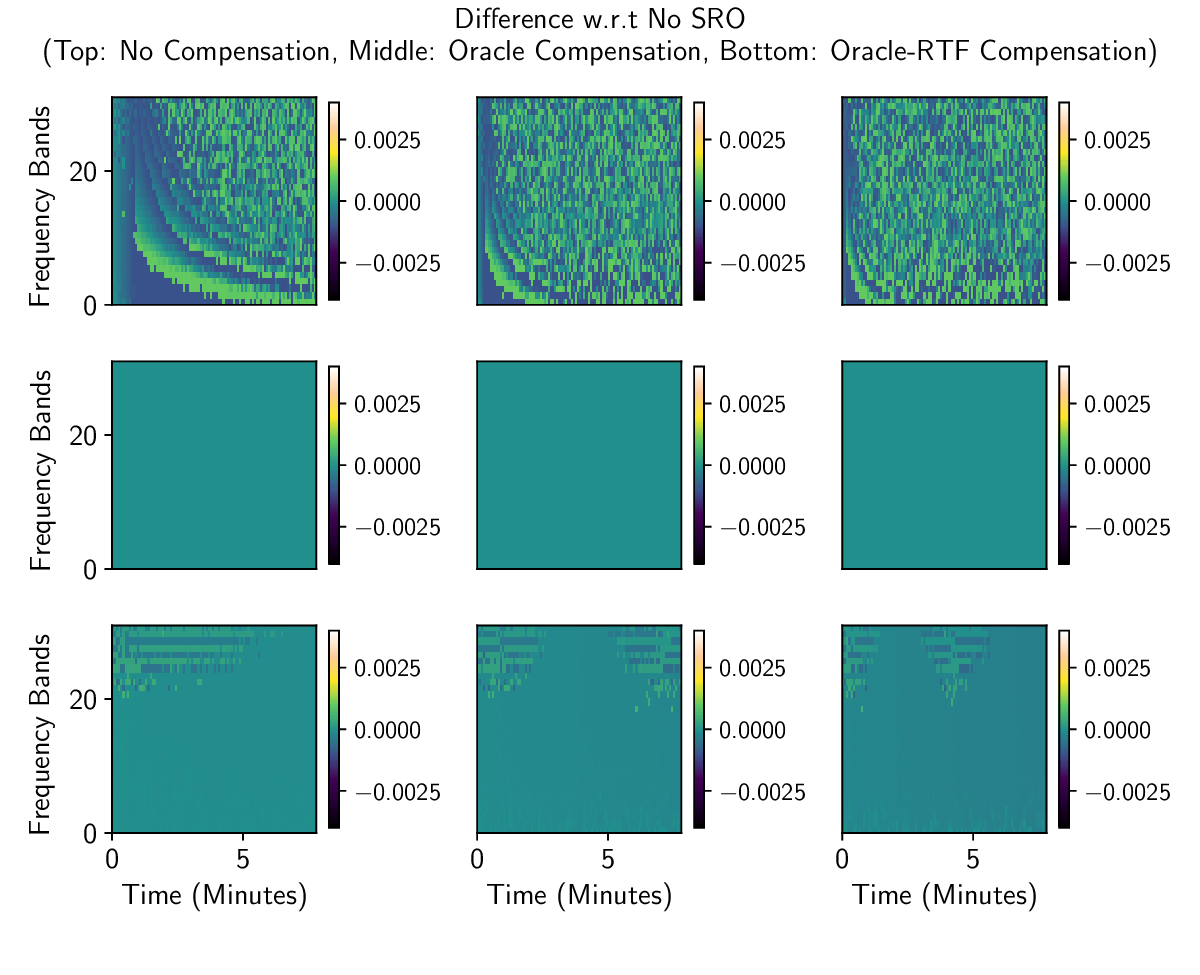}
  }
  \hfill
  \subfigure[IC]{
    \includegraphics[width=0.485\linewidth]{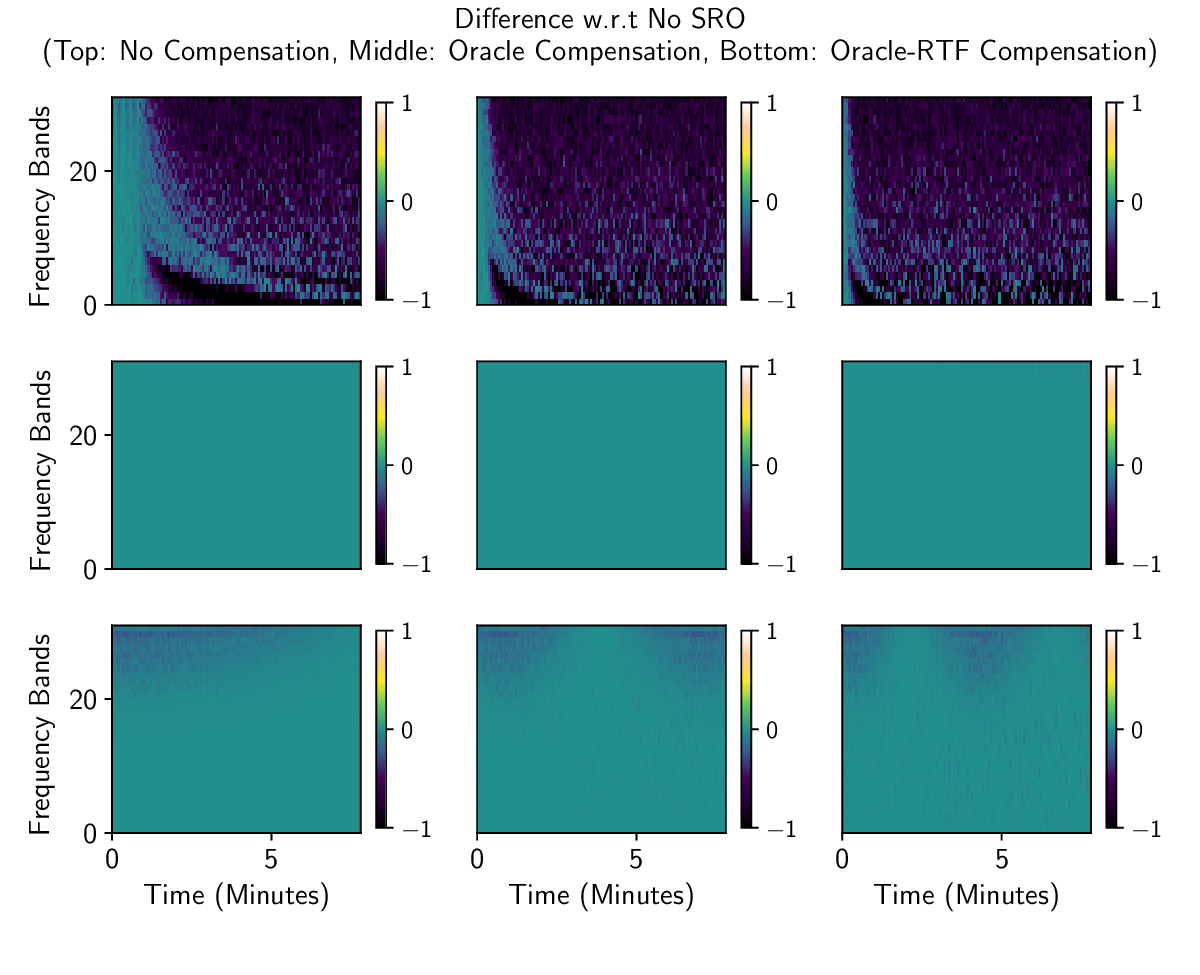}
  }
  \caption{Difference plots of \ac{ITD} and \ac{IC} w.r.t. no-SRO for the conditions (\,top: no compensation, middle: oracle compensation, bottom: oracle-\ac{RTF}-based compensation\,) at three different \ac{SRO} configuration (\,left: [10,\,-10]\,\ac{ppm},\,mid: [10,\,-50]\,\ac{ppm},\,right: [10,\,-100]\,\ac{ppm}\,).}
  \label{fig:objective_metrics}
\end{figure*}

To estimate the~\ac{SRO}, we use the~\ac{DWACD} algorithm proposed in~\cite{gburrek_synchronization_2022}. Given the input signal $\widehat{Z}_q\,\left[k,l\right]$ and the reference signal $X_q\,\left[k,l\right]$, the \ac{SRO} estimation relies on first computing the complex coherence function $\Gamma\left[k,l\right]$, which is given by
\begin{equation}
\label{coherence_func_ACD}
\Gamma\left[k,l\right] = \frac{\Phi_{\widehat{Z}_q\,X_q}\left[k,l\right]}{\sqrt{\Phi_{\widehat{Z}_q\,\widehat{Z}_q}\left[k,l\right]\Phi_{X_q\,X_q}\left[k,l\right]}}, 
\end{equation}
where $\Phi_{\widehat{Z}_q\,X_q}$, $\Phi_{\widehat{Z}_q\,\widehat{Z}_q}$ and $\Phi_{X_q\,X_q}$ are the cross and auto \acp{PSD}, respectively. The phase function $\widetilde{P}\left[k,l\right]$ is computed by the complex conjugate product of two consecutive complex coherence functions with a temporal distance of $L$, i.e., 
\begin{equation}
\label{avg_coherence_func_wACD}
\widetilde{P}\left[k,l\right] = \Gamma\left[k,l+L\right]~\Gamma^*\left[k,l\right].
\end{equation}
The temporally averaged phase function $P\left[k,l\right]$ and the \ac{GCC} $p(\beta,l)$ are given by
\begin{equation}
\label{avg_coherence_func_dwACD}
P\left[k,l\right] = \alpha_s~P\left[k,l-1\right] + (1-\alpha_s)~\widetilde{P}\left[k,l\right],
\end{equation}
\begin{equation}
\label{equ:gcc}
p\left[\beta,l\right] = \textrm{IDFT}\{P\left[k,l\right]\},
\end{equation}
where $\alpha_s$ is the smoothing factor, $\beta$ is the time-lag, and $\textrm{IDFT}$ is the inverse DFT. 
The estimated~\ac{SRO} $\widehat{\bar{\epsilon}}_{q}$ is obtained by first finding the integer time-lag $\beta_\text{max}$ that maximizes the~\ac{GCC}, i.e., 
\begin{equation}
\label{SRO_estimate_dwACD_1}
\widehat{\bar{\epsilon}}_{q}\left[l\right] = -\frac{1}{L\,N_{\text{h}}}\beta_\text{max} = -\frac{1}{L\,N_{\text{h}}}\,\arg \max_{\beta}|p\left[\beta,l\right]|.
\end{equation} 
Then, an \ac{SRO} estimate is obtained by determining the non-integer time-lag by performing a golden search in the interval given by $[\beta_\text{max}-0.5,\beta_\text{max}+0.5]$. 
The complex coherence function $\Gamma\left[k,l\right]$ is estimated only when signal activity is detected in both signals $\widehat{Z}_q\left[k,l\right]$ and $X_q\left[k,l\right]$. In this study, we used the energy-based threshold to detect signal activity. To avoid temporal fluctuations in the estimated \ac{SRO}, we apply temporal smoothing on the estimated \ac{SRO}.  

\section{Experimental Results}

\begin{figure}[t]
    \centering
    \includegraphics[width=\columnwidth]{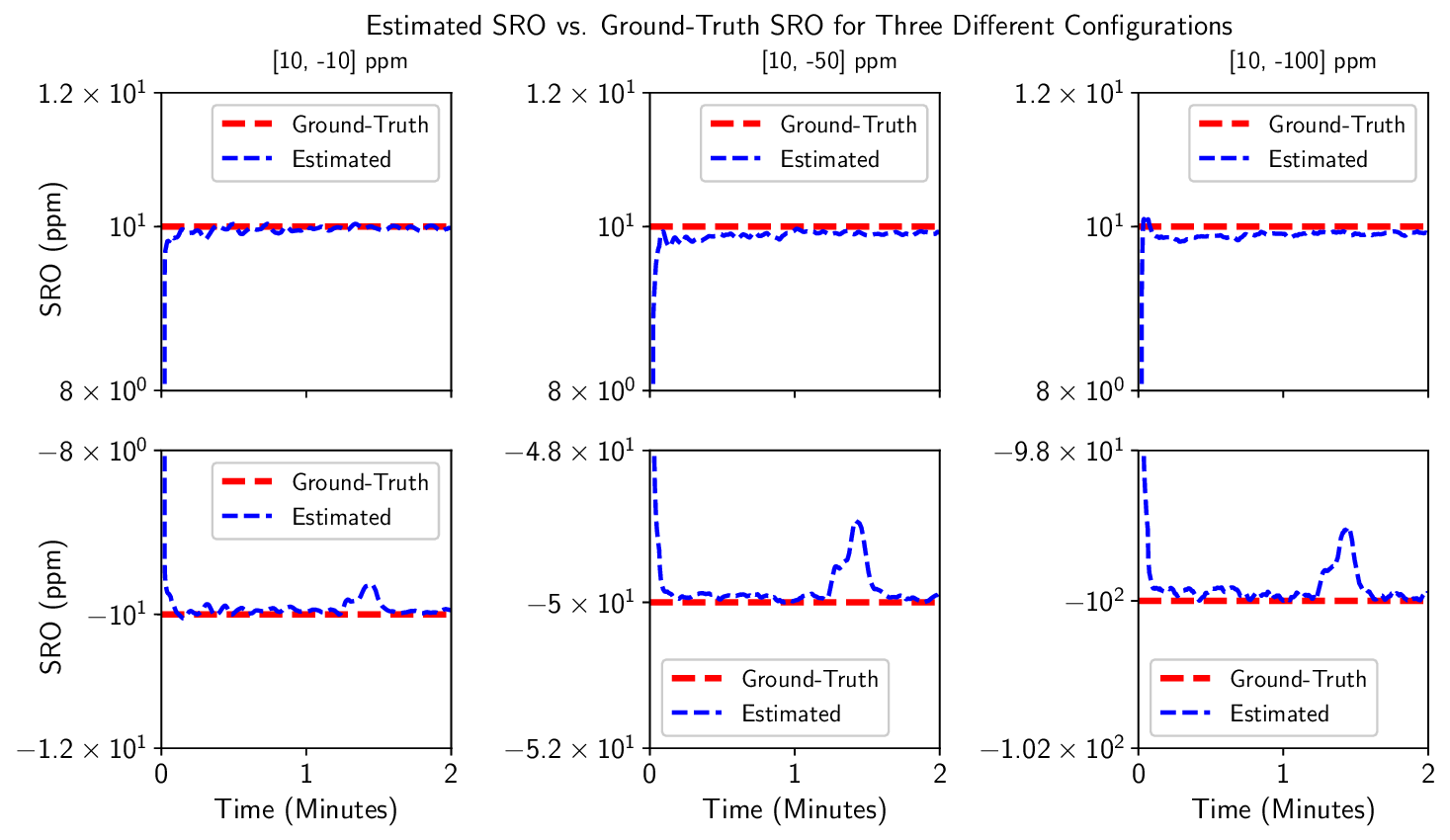}
    \caption{Estimated \ac{SRO} in \ac{ppm} (top: $\widehat{\bar{\epsilon}}_1$, bottom: $\widehat{\bar{\epsilon}}_2$) vs. Ground-Truth \ac{SRO} for three different \ac{SRO} configuration (\,left: [10,\,-10]\,\ac{ppm},\,mid: [10,\,-50]\,\ac{ppm},\,right: [10,\,-100]\,\ac{ppm}\,) for the first two minutes.}
    \label{fig:sro_estimation}
\end{figure}


For our evaluation, we considered a room of size [7,\,7,\,6]\,\unit{m} with an RT60 value of 0.3\,\unit{s}. 
Each channel of the stereo playback signal was considered to be an omnidirectional source. These sources were placed at [2.2,\,3.4,\,1.8]\,\unit{m} and [5.2,\,3.5,\,2.1]\,\unit{m}, respectively. The primary device consisted of a circular microphone array with $M=4$ microphones centered at [3.75,\,3.35,\,2.0]\,\unit{m} and a radius of 10\,\unit{cm}. With these parameters, the microphone signals were generated at a sampling rate of 16\,\unit{kHz} using Pyroomacoustics \cite{Scheibler_2018}. We simulated three \ac{SRO} configurations $(\epsilon_1,\,\epsilon_2)$: (10,\,-10)\,\unit{ppm}, (10,\,-50)\,\unit{ppm} and (10,\,-100)\,\unit{ppm} on the microphone signal using the STFT method proposed in~\cite{Schmalenstroeer_overlapsave_resampling_eusipco_2018} using a segment length of 8192 samples~\cite{gburrek_synchronization_2022}. The smoothing factors $\alpha_s$, $\alpha$ and the hop-size $N_\text{h}$ are set to 0.95, $1 \times 10^{-6}$ and 2048, respectively. 

Figure~\,\ref{fig:objective_metrics} shows the difference plots with the binaural cues \acf{ITD} and \acf{IC} w.r.t. no-\ac{SRO} for the following conditions: no compensation, oracle compensation, oracle-\ac{RTF}-based compensation. The \acp{ITD} were computed using the model\,\cite{May_2011} implemented in the auditory modeling toolbox\,\cite{majdak_amt_2022} for a synthetic stereo signal of length 8\,\unit{minutes} containing the same Gaussian white noise in both channels. When the \ac{SRO} is not compensated, we can see that the coherence reaches the value of zero (i.e., the coherence decreases) faster at higher \acp{SRO} compared to lower \acp{SRO}. Also, the model prediction of \ac{ITD} is affected by the \ac{SRO}. When the \ac{SRO} is perfectly compensated, we see that the effect of \ac{SRO} on binaural cues is perfectly compensated. Oracle-\ac{RTF} based compensation shows that the effect of \ac{SRO} on the binaural cues can be compensated, especially at low and mid frequencies. At higher frequencies, the effect of \ac{SRO} on the binaural cues is minimized but not perfectly compensated.    

Figure~\,\ref{fig:sro_estimation} shows the plots for the estimated \ac{SRO} for the first two minutes of the eight-minute file computed by averaging over seven different files. From the plots, it can be concluded that the estimated \ac{SRO} is highly robust.   

\begin{figure}[t]
    \centering
    \includegraphics[width=\columnwidth]{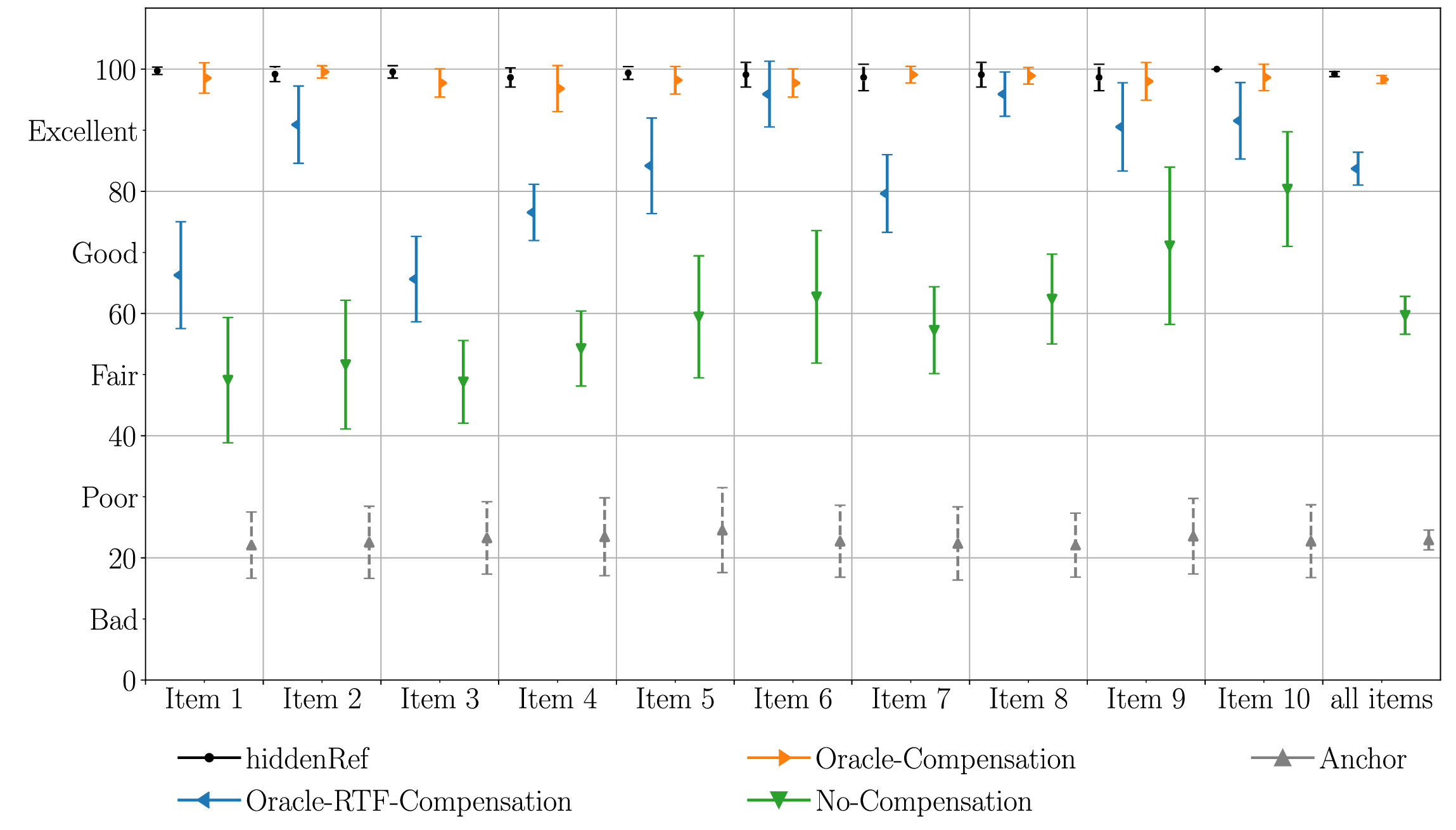}
    \caption{MUSHRA listening test results for 11 listeners.}
    \label{fig:subjective_test}
\end{figure}

A subjective listening test using stereo signals (Rock, Instrumental, and Classical Singing) was conducted according to the \ac{MUSHRA} methodology,\,\cite{ITU-R-BS.1534-3}, under two \ac{SRO} configurations ([10,\,-10]\,\unit{ppm} and [10,\,-100]\,\unit{ppm}). The test involved 11 listeners and included the following conditions: hidden reference (black), ground-truth compensation (orange), oracle-\ac{RTF}-based compensation (blue), no compensation (green), and anchor (grey). To enable headphone-based evaluation, the signals were binauralized using impulse responses corresponding to loudspeakers placed at $\pm 90\unit{deg}$,\cite{Pike2017}. For the test, 25\,\unit{s} segments were extracted around the 6\textsuperscript{th} minute from 8-minute-long audio files. The anchor was created by first computing a passive downmix, followed by low-pass filtering with a 3.5\,\unit{kHz} cutoff. Figure~\ref{fig:subjective_test} shows the \ac{MUSHRA} results. The results indicate that \ac{SRO} affects listener perception. The proposed oracle-\ac{RTF}-based compensation significantly reduces this effect, though it does not eliminate it entirely.  

\section{Conclusion}
We proposed an audio-domain \ac{SRO} compensation method that mitigates the impact of \acp{SRO} on spatial audio reproduction in wireless audio transmission scenarios. Factors other than \acp{SRO}, such as latency and audio coding — common in wireless audio reproduction systems — were beyond the scope of this study and are considered future work. Evaluation using both objective metrics and subjective listening tests shows that the proposed method effectively preserves binaural cues and reduces perceptual degradation caused by \acp{SRO}.
 
\IEEEtriggeratref{18}
\bibliographystyle{IEEEtran}
\bibliography{refs}

\begin{thebibliography}{10}
\providecommand{\url}[1]{#1}
\csname url@samestyle\endcsname
\providecommand{\newblock}{\relax}
\providecommand{\bibinfo}[2]{#2}
\providecommand{\BIBentrySTDinterwordspacing}{\spaceskip=0pt\relax}
\providecommand{\BIBentryALTinterwordstretchfactor}{4}
\providecommand{\BIBentryALTinterwordspacing}{\spaceskip=\fontdimen2\font plus
\BIBentryALTinterwordstretchfactor\fontdimen3\font minus \fontdimen4\font\relax}
\providecommand{\BIBforeignlanguage}[2]{{%
\expandafter\ifx\csname l@#1\endcsname\relax
\typeout{** WARNING: IEEEtran.bst: No hyphenation pattern has been}%
\typeout{** loaded for the language `#1'. Using the pattern for}%
\typeout{** the default language instead.}%
\else
\language=\csname l@#1\endcsname
\fi
#2}}
\providecommand{\BIBdecl}{\relax}
\BIBdecl

\bibitem{Rumsey_2001}
\BIBentryALTinterwordspacing
F.~Rumsey, \emph{Spatial Audio}, ser. Music Technology Series.\hskip 1em plus 0.5em minus 0.4em\relax Taylor \& Francis, 2001. [Online]. Available: \url{https://www.taylorfrancis.com/books/mono/10.4324/9780080498195/spatial-audio-francis-rumsey}
\BIBentrySTDinterwordspacing

\bibitem{Roginska_2017}
\BIBentryALTinterwordspacing
A.~Roginska and P.~Geluso, Eds., \emph{Immersive Sound: The Art and Science of Binaural and Multi-Channel Audio}, ser. Audio Engineering Society Presents.\hskip 1em plus 0.5em minus 0.4em\relax Routledge, 2017. [Online]. Available: \url{https://www.taylorfrancis.com/books/edit/10.4324/9781315707525/immersive-sound-agnieszka-roginska-paul-geluso}
\BIBentrySTDinterwordspacing

\bibitem{ieee1588-2008}
\BIBentryALTinterwordspacing
{IEEE Instrumentation and Measurement Society}, \emph{{IEEE Standard for a Precision Clock Synchronization Protocol for Networked Measurement and Control Systems}}, Institute of Electrical and Electronics Engineers Std. IEEE 1588-2008, July 2008. [Online]. Available: \url{https://standards.ieee.org/standard/1588-2008.html}
\BIBentrySTDinterwordspacing

\bibitem{rfc5905}
\BIBentryALTinterwordspacing
D.~Mills, J.~Martin, J.~Burbank, and W.~Kasch, ``{Network Time Protocol Version 4: Protocol and Algorithms Specification},'' Request for Comments 5905, June 2010. [Online]. Available: \url{https://datatracker.ietf.org/doc/html/rfc5905}
\BIBentrySTDinterwordspacing

\bibitem{Culbert_2006}
\BIBentryALTinterwordspacing
M.~Culbert and A.~Lindahl, ``Method and system for time synchronizing multiple loudspeakers,'' Patent US20\,060\,067\,536A1, March 30, 2006. [Online]. Available: \url{https://patents.google.com/patent/US20060067536A1/en}
\BIBentrySTDinterwordspacing

\bibitem{Lauri2015}
\BIBentryALTinterwordspacing
C.~Lauri and J.~Malmgren, ``{Synchronization of Streamed Audio Between Multiple Playback Devices Over an Unmanaged IP Network},'' Master's thesis, Lund University, October 2015. [Online]. Available: \url{https://lup.lub.lu.se/student-papers/record/8052964/file/8052965.pdf}
\BIBentrySTDinterwordspacing

\bibitem{SMG_BSRO_2012}
S.~Markovich-Golan, S.~Gannot, and I.~Cohen, ``Blind sampling rate offset estimation and compensation in wireless acoustic sensor networks with application to beamforming,'' in \emph{International Workshop on Acoustic Signal Enhancement}, 2012, pp. 1--4.

\bibitem{wang_CMSRO_WASN_2016}
L.~Wang and S.~Doclo, ``Correlation maximization-based sampling rate offset estimation for distributed microphone arrays,'' \emph{IEEE/ACM Transactions on Audio, Speech, and Language Processing}, vol.~24, no.~3, pp. 571--582, 2016.

\bibitem{schmalenstroeer_multi-stage_2017}
\BIBentryALTinterwordspacing
J.~Schmalenstroeer, J.~Heymann, L.~Drude, C.~Boeddecker, and R.~Haeb-Umbach, ``\BIBforeignlanguage{en}{Multi-stage coherence drift based sampling rate synchronization for acoustic beamforming},'' in \emph{\BIBforeignlanguage{en}{{IEEE} 19th {International} {Workshop} on {Multimedia} {Signal} {Processing} ({MMSP})}}.\hskip 1em plus 0.5em minus 0.4em\relax Luton: IEEE, Oct. 2017, pp. 1--6. [Online]. Available: \url{http://ieeexplore.ieee.org/document/8122278/}
\BIBentrySTDinterwordspacing

\bibitem{chinaev_online_2021}
\BIBentryALTinterwordspacing
A.~Chinaev, G.~Enzner, T.~Gburrek, and J.~Schmalenstroeer, ``\BIBforeignlanguage{en}{Online {Estimation} of {Sampling} {Rate} {Offsets} in {Wireless} {Acoustic} {Sensor} {Networks} with {Packet} {Loss}},'' in \emph{\BIBforeignlanguage{en}{29th {European} {Signal} {Processing} {Conference} ({EUSIPCO})}}.\hskip 1em plus 0.5em minus 0.4em\relax Dublin, Ireland: IEEE, Aug. 2021, pp. 1110--1114. [Online]. Available: \url{https://ieeexplore.ieee.org/document/9616037/}
\BIBentrySTDinterwordspacing

\bibitem{gburrek_synchronization_2022}
\BIBentryALTinterwordspacing
T.~Gburrek, J.~Schmalenstroeer, and R.~Haeb-Umbach, ``\BIBforeignlanguage{en}{On {Synchronization} of {Wireless} {Acoustic} {Sensor} {Networks} in the {Presence} of {Time}-{Varying} {Sampling} {Rate} {Offsets} and {Speaker} {Changes}},'' in \emph{\BIBforeignlanguage{en}{{IEEE} {International} {Conference} on {Acoustics}, {Speech} and {Signal} {Processing} ({ICASSP})}}.\hskip 1em plus 0.5em minus 0.4em\relax Singapore, Singapore: IEEE, May 2022, pp. 916--920. [Online]. Available: \url{https://ieeexplore.ieee.org/document/9746284/}
\BIBentrySTDinterwordspacing

\bibitem{pawig_adaptive_2010}
\BIBentryALTinterwordspacing
M.~Pawig, G.~Enzner, and P.~Vary, ``\BIBforeignlanguage{en}{Adaptive {Sampling} {Rate} {Correction} for {Acoustic} {Echo} {Control} in {Voice}-{Over}-{IP}},'' \emph{\BIBforeignlanguage{en}{IEEE Transactions on Signal Processing}}, vol.~58, no.~1, pp. 189--199, Jan. 2010. [Online]. Available: \url{http://ieeexplore.ieee.org/document/5170064/}
\BIBentrySTDinterwordspacing

\bibitem{abe_frequency_2014}
\BIBentryALTinterwordspacing
M.~Abe and M.~Nishiguchi, ``\BIBforeignlanguage{en}{Frequency domain acoustic echo canceller that handles asynchronous {A}/{D} and {D}/{A} clocks},'' in \emph{\BIBforeignlanguage{en}{{IEEE} {International} {Conference} on {Acoustics}, {Speech} and {Signal} {Processing} ({ICASSP})}}.\hskip 1em plus 0.5em minus 0.4em\relax Florence, Italy: IEEE, May 2014, pp. 5924--5928. [Online]. Available: \url{http://ieeexplore.ieee.org/document/6854740/}
\BIBentrySTDinterwordspacing

\bibitem{helwani_clock_2022}
\BIBentryALTinterwordspacing
K.~Helwani, E.~Soltanmohammadi, M.~M. Goodwin, and A.~Krishnaswamy, ``\BIBforeignlanguage{en}{Clock {Skew} {Robust} {Acoustic} {Echo} {Cancellation}},'' in \emph{\BIBforeignlanguage{en}{Interspeech 2022}}.\hskip 1em plus 0.5em minus 0.4em\relax ISCA, Sep. 2022, pp. 2533--2537. [Online]. Available: \url{https://www.isca-speech.org/archive/interspeech_2022/helwani22_interspeech.html}
\BIBentrySTDinterwordspacing

\bibitem{korse_2024}
S.~Korse, O.~Thiergart, and E.~A.~P. Habets, ``{Sample Rate Offset Compensated Acoustic Echo Cancellation for Multi-Device Scenarios},'' in \emph{18th International Workshop on Acoustic Signal Enhancement (IWAENC)}, 2024, pp. 1--5.

\bibitem{VanVeen_1988}
\BIBentryALTinterwordspacing
B.~D.~V. Veen and K.~M. Buckley, ``Beamforming: A versatile approach to spatial filtering,'' \emph{IEEE ASSP Magazine}, vol.~5, no.~2, pp. 4--24, 1988. [Online]. Available: \url{https://ieeexplore.ieee.org/document/665}
\BIBentrySTDinterwordspacing

\bibitem{VanTrees2002}
\BIBentryALTinterwordspacing
H.~L.~V. Trees, \emph{Optimum Array Processing: Part IV of Detection, Estimation, and Modulation Theory}.\hskip 1em plus 0.5em minus 0.4em\relax Hoboken, NJ, USA: Wiley-Interscience, 2002. [Online]. Available: \url{https://onlinelibrary.wiley.com/doi/book/10.1002/0471221104}
\BIBentrySTDinterwordspacing

\bibitem{Frost1972}
\BIBentryALTinterwordspacing
O.~L.~F. III, ``An algorithm for linearly constrained adaptive array processing,'' \emph{Proceedings of the IEEE}, vol.~60, no.~8, pp. 926--935, August 1972. [Online]. Available: \url{https://ieeexplore.ieee.org/document/1451993}
\BIBentrySTDinterwordspacing

\bibitem{chakrabarty2016numerical}
\BIBentryALTinterwordspacing
S.~Chakrabarty and E.~A.~P. Habets, ``On the numerical instability of an lcmv beamformer for a uniform linear array,'' \emph{IEEE Signal Processing Letters}, vol.~23, no.~2, pp. 272--276, Feb 2016. [Online]. Available: \url{https://ieeexplore.ieee.org/document/7355407}
\BIBentrySTDinterwordspacing

\bibitem{bishop2006pattern}
\BIBentryALTinterwordspacing
C.~M. Bishop, \emph{Pattern Recognition and Machine Learning}.\hskip 1em plus 0.5em minus 0.4em\relax New York, NY, USA: Springer, 2006. [Online]. Available: \url{https://books.google.com/books/about/Pattern_Recognition_and_Machine_Learning.html?id=kTNoQgAACAAJ}
\BIBentrySTDinterwordspacing

\bibitem{Chen2008}
\BIBentryALTinterwordspacing
J.~Chen, J.~Benesty, and Y.~Huang, ``A minimum distortion noise reduction algorithm with multiple microphones,'' \emph{IEEE Transactions on Audio, Speech, and Language Processing}, vol.~16, no.~3, pp. 481--493, March 2008. [Online]. Available: \url{https://ieeexplore.ieee.org/document/4431805}
\BIBentrySTDinterwordspacing

\bibitem{Markovich2009}
\BIBentryALTinterwordspacing
S.~Markovich, S.~Gannot, and I.~Cohen, ``Multichannel eigenspace beamforming in a reverberant noisy environment with multiple interfering speech signals,'' \emph{IEEE Transactions on Audio, Speech, and Language Processing}, vol.~17, no.~6, pp. 1071--1086, August 2009. [Online]. Available: \url{https://ieeexplore.ieee.org/document/4814760}
\BIBentrySTDinterwordspacing

\bibitem{Taseska_2015}
\BIBentryALTinterwordspacing
M.~Taseska and E.~A.~P. Habets, ``Relative transfer function estimation exploiting instantaneous signals and the signal subspace,'' in \emph{Proceedings of the 23rd European Signal Processing Conference (EUSIPCO)}, Nice, France, August 2015, pp. 404--408. [Online]. Available: \url{https://ieeexplore.ieee.org/document/7362414}
\BIBentrySTDinterwordspacing

\bibitem{Scheibler_2018}
\BIBentryALTinterwordspacing
R.~Scheibler, E.~Bezzam, and I.~Dokmanić, ``Pyroomacoustics: A python package for audio room simulation and array processing algorithms,'' in \emph{Proceedings of the IEEE International Conference on Acoustics, Speech and Signal Processing (ICASSP)}, 2018, pp. 351--355. [Online]. Available: \url{https://ieeexplore.ieee.org/document/8461310}
\BIBentrySTDinterwordspacing

\bibitem{Schmalenstroeer_overlapsave_resampling_eusipco_2018}
J.~Schmalenstroeer and R.~Haeb-Umbach, ``Efficient sampling rate offset compensation - an overlap-save based approach,'' in \emph{26th European Signal Processing Conference (EUSIPCO)}, 2018, pp. 499--503.

\bibitem{May_2011}
T.~May, S.~van~de Par, and A.~Kohlrausch, ``A probabilistic model for robust localization based on a binaural auditory front-end,'' \emph{IEEE Transactions on Audio, Speech, and Language Processing}, vol.~19, no.~1, pp. 1--13, 2011.

\bibitem{majdak_amt_2022}
\BIBentryALTinterwordspacing
{Majdak, Piotr}, {Hollomey, Clara}, and {Baumgartner, Robert}, ``Amt 1.x: A toolbox for reproducible research in auditory modeling,'' \emph{Acta Acust.}, vol.~6, p.~19, 2022. [Online]. Available: \url{https://doi.org/10.1051/aacus/2022011}
\BIBentrySTDinterwordspacing

\bibitem{ITU-R-BS.1534-3}
\BIBentryALTinterwordspacing
{International Telecommunication Union}, \emph{{ITU-R BS.1534-3: Method for the subjective assessment of intermediate quality level of audio systems}}, Std. ITU-R BS.1534-3, October 2015. [Online]. Available: \url{https://www.itu.int/rec/R-REC-BS.1534-3-201510-I/en}
\BIBentrySTDinterwordspacing

\bibitem{Pike2017}
\BIBentryALTinterwordspacing
C.~Pike and M.~Romanov, ``An impulse response dataset for dynamic data-based auralisation of advanced sound systems,'' in \emph{Proceedings of the 142nd Audio Engineering Society Convention}, Berlin, Germany, May 2017, engineering Brief 334. [Online]. Available: \url{https://www.aes.org/e-lib/browse.cfm?elib=18709}
\BIBentrySTDinterwordspacing

\end{thebibliography}

\end{document}